# Grain boundary metastability controls irradiation resistance in nanocrystalline metals

Osman El-Atwani[1,2], Annie K. Barnett[6], Enrique Martínez[12],Jian Han[3,4], Asher C. Leff[2,5], Chang-Yu Hung[6], James E. Nathaniel[2,7], Sicong He[8], Emily H. Mang[6], Larissa M. Woryk[4], Khalid Hattar[9,10], Blas P. Uberuaga[11], David J. Srolovitz[4,13,14], Michael L. Falk[6], Jaime Marian[9], and Mitra L. Taheri[2,6]*

[1]Energy and Environment Division, Pacific Northwest National Laboratory, Richland, WA

[2]Dept. of Materials Science & Engineering, Drexel University, Philadelphia, PA

[3]Dept. of Materials Science & Engineering, City University of Hong Kong, Hong Kong SAR, CHINA

[4]Dept. of Materials Science & Engineering, University of Pennsylvania, Philadelphia, PA

[5]Army Research Laboratories, Adelphi, MD

[6]Dept. of Materials Science & Engineering, Johns Hopkins University, Baltimore, MD

[7]Sandia National Laboratories, Livermore, CA

[8]University of California Los Angeles, Los Angeles, CA

[9]Center for Integrated Nanotechnologies, Sandia National Laboratories, Albuquerque, NM

[10] Department of Nuclear Engineering, University of Tennessee, Knoxville, TN

[11]Materials Science and Technology Division, Los Alamos National Laboratory, Los Alamos, NM

[12]Dept. of Mechanical Engineering, Clemson University, Clemson, SC 29623 USA

[13]Dept.of Mechanical Engineering, The University of Hong Kong, Hong Kong SAR, CHINA

[14]International Digital Economy Academy (IDEA), Shenzhen, CHINA

* M.L. Taheri

**Email:** mtaheri4@jhu.edu

**Abstract**
Grain boundaries (GBs) in polycrystalline materials are powerful sinks for irradiation defects. While standard theories assume that a GB's efficiency as a sink is defined solely by its character before irradiation, recent evidence conclusively shows that the irradiation sink efficiency is a highly dynamic property controlled by the intrinsic metastability of GBs under far-from-equilibrium irradiation conditions. In this paper, we reveal that the denuded (i.e., defect-free) zone, typically the signature of a strong sink, can collapse as irradiation damage accumulates. We propose a radiation damage evolution model that captures this behavior based on the emergence of a series of irradiation defect-enabled metastable GB microstate changes that dynamically alter the ability of the GB to absorb further damage. We show that these microstate changes control further defect absorption and give rise to the formation of a defect network that manifests itself as a net Nye-tensor signal detectable via lattice curvature experiments.

**Teaser**

Experimental and simulation results elucidate outstanding questions on how nanocrystalline materials respond to irradiation extremes.

**Introduction**

Controlling microstructural evolution to tailor properties, such as radiation damage tolerance [1–29], is a key goal of materials design [29]. Under irradiation, materials are driven far from equilibrium through a sequence of transient states that determine the ability of a material to continue to accommodate damage without losing integrity. As such, harnessing this metastability is key to understanding properties and routes to effect damage accommodation brought about by the evolving microstructure. Crystalline interfaces or grain boundaries present an opportunity for tuning non-equilibrium states. Indeed, grain boundaries exhibit metastable structures during irradiation [30–32]. Understanding and controlling nonequilibrium interfacial behavior represents a critical step forward for materials innovation in a wide variety of applications, including nuclear power [2,20,33], catalysis [34,35], and thermoelectric performance [36]. Future materials will be designed and optimized to account for otherwise "unseen" grain boundary structures for realizing key outcomes, such as damage tolerance.

To date, no material is known to be fully immune to radiation damage, especially beyond a dose level of several hundred displacements per atom[1] (dpa). Although a large fraction of the point defects produced by damage cascades recombine immediately following a damage event, the residual defects can lead to the accumulation of various forms of radiation damage and the evolution of the microstructure, ultimately leading to failure. Among various proposed methods [6,17,18] for achieving radiation damage resistance, the adoption of predesigned sinks to reduce or eliminate radiation-induced defects has become particularly promising [20,37,38]. Specifically, designing a microstructure with an increased density of specific types of grain boundaries (GBs) is known to benefit a wide range of properties, such as corrosion resistance [28] or fatigue resistance [10]. A microstructure with a predetermined distribution of specific GB types (GB engineering) may promote recombination or annihilation of irradiation-induced point defects at GBs and also control segregation of elements critical to the chemical and mechanical stability of structural alloys [2,21].

GBs act as sinks for point defects during irradiation and hence modify the distribution of point and other defects (dislocations, stacking fault tetrahedra, etc.) within the grains. If the rate of defect annihilation at the boundary is high relative to the defect generation rate, then a (relatively) defect-free, or denuded, zone can form adjacent to the boundary. This phenomenon has been thoroughly studied in literature [9, 37, 78-81] and aids in understanding the response of materials under irradiation. The width of the denuded zone is proportional to the sink efficiency of the boundary and the rate of defect generation (a function of the incident particle type, flux, and energy, and temperature). However, this relationship is not straightforward and the GB sink efficiency may be related to both point defect mobility and disconnection dynamics within the boundary, as well as processes in the grain interior, such as defect recombination within the grains [26,39]. Each point defect absorption event inherently changes the structure of the GB (new atoms are being added to/removed from the structure. If line defects, such as those created from the accumulation of point defects (e.g., dislocation loops) are absorbed at GBs, they too modify the GB structure, either by interacting with the native disconnections (line defects characterized by a Burgers vector and a step height that give GBs their character) that form the intrinsic GB structure or forming extrinsic disconnections that alter the microscopic GB character and/or the macroscopic GB bi-crystallography. The formation of extrinsic defects could have significant effects on GB structure and point defect mobility within the GB, and hence, its sink efficiency. The sink efficiency, $\eta$, is a measure of the ability of an interface to absorb point-defects [19,33,35–37]; it is frequently defined as the ratio of the actual point defect flux into the interface to that for a perfect sink, $\eta = J_{\text{measure}}/J_{\text{perfect}}$. Therefore, an interface with high $\eta$ can remove defects from its surroundings at a higher rate than one with low $\eta$. Beyerlein et al. [37] proposed the following relationship between denuded zone width $\lambda$ and sink efficiency

$$\lambda_{\text{v,i}} \sqrt{\frac{K_{\text{sv,i}}}{D_{\text{v,i}}}} = \ln \eta_{\text{v,i}} - \ln\left(1 - c^*_{\text{v,i}} \frac{K_{\text{sv,i}}}{K_0}\right), \qquad (1)$$

where $\lambda_{\text{v,i}}$ is the width of the denuded zone, $K_{\text{sv,i}}$ is the defect-sink reaction rate coefficient, $K_0$ is the defect production rate, $D_{\text{v,i}}$ is the defect diffusivity within the grain, $c^*_{\text{v,i}}$ is the critical defect concentration to nucleate loops or voids, and $\eta_{\text{v,i}}$ is the sink efficiency describing the ability of the sink to absorb point defects; the subscript "v" denotes vacancies and "i" denotes interstitials. $\eta$ may be between 0 (no defect absorption) and 1 (perfect sink) and is an overall GB sink figure of merit.

---

[1] Equivalent to ~100 years in existing light water reactors, but only 10-20 in Generation IV designs, and 1-2 in fusion reactor designs.

In general, no solid/solid interfaces exist which are *perfect* sinks. Although earlier studies treated GBs as perfect sinks, recent works [2,7–9,26,37,42–44] demonstrate that some GBs more effectively absorb defects than others. For example, experimental observations [6,8,42,45] have shown a dependence of denuded zone widths on GB type, although these trends do not reveal a consistent relation and can thus only be viewed as suggestive. As denuded zone width is a direct consequence of sink efficiency, we infer that varying denuded zone widths imply varying sink efficiency (at fixed thermodynamic and irradiation conditions). Despite a history of research linking GB structure to its effect on properties (such as radiation-induced segregation) [3], defect absorption rates do not follow a clear trend with macroscopic GB descriptors (misorientation, inclination) [4,7–9,27]. Theories were proposed to explain these variations, including local tensile strain [4,15,16], GB stability, GB free volume [26], and GB microstates [25], but no unifying theory has been able to explain the variation in sink efficiency even for similar interfaces or GB types. Recent simulations [8,19] that classify GBs according to microstates, or phases, present an opportunity to improve our understanding of sink efficiency, to explain the resulting denuded zone variations, and to glean a predictive understanding of interfaces to realize radiation tolerant microstructures. GB stability may be defined as the ability of a GB to continue to absorb point defects without becoming saturated and without changing its macroscopic degrees of freedom (DOFs), e.g., misorientation and inclination. To maintain stability, we hypothesize that GBs evolve via metastable microstates as point defects are added to the boundary [12]. The idea of metastable GB states has been discussed over many years [1,5,12,14,22,25], and while it has been shown that adding or removing atoms can change the GB microstate [11,12,19,23,24], the impact of a GB microstate change on the sink efficiency, or, importantly, how irradiation itself leads to evolution of the micro-states, remains unanswered [13,25,26].

Most GBs exhibit multiple metastable structural states (even in pure materials). While the number of such states tends to be small in systems with high-symmetry GBs (i.e., coherent twins), the number may be large for most other GBs [1,12,22,23,46,47] (even many with small coincident site densities, Σ). Different states can emerge when different regions/domains of the GB may be in different structural states separated by line defects (disconnections, GB dislocations, partial GB dislocations, etc.) [11,12,48]. We assert that when a GB experiences a microstate change, it no longer possesses the same absorption efficiency for defects [26]. This could lead to several events, including defect pileups, saturation of defects in the GB region and change in denuded zone width.

Through a combination of quantitative *in situ* electron microscopy and atomistic modeling, we define a relationship between GB sink efficiency limits and the local defect landscape created by the irradiation dose. GBs are revealed to "shut down" as sinks at particular doses of irradiation, as evidenced by denuded zone collapse and a related mesoscale "signature" of dislocation density within the GB. Interestingly, the change in sink efficiency, and increased dislocation density in the GB region due to denuded zone collapse, contributes to a complex defect network adjacent to the GB. While we connect the change in sink efficiency to microstate changes in the GB, we note that denuded zone collapse may, of course, be enhanced by local defect bias, accelerating changes in the GB state (supporting simulations added in the *Supplementary Information*). Critically, this finding reveals an ability to tune GB response not only by their innate structure or state, but also by external factors that control local defect environments.

**Results**

We begin with a series of striking experimental observations that emanate from our *in situ* 10 keV-$He^+$ irradiation TEM experiments of polycrystalline Fe films (see *Methods* for experimental details). While denuded zone evolution was found to be GB-dependent, with denuded zones developed at different $He^+$ doses around different GBs, a particularly remarkable observation was that the denuded zone width evolves with time (dose) (in agreement with the observations by Yang et al. in a different material system [49]), and several denuded zone collapse events are seen to occur (**Figure. 1**).

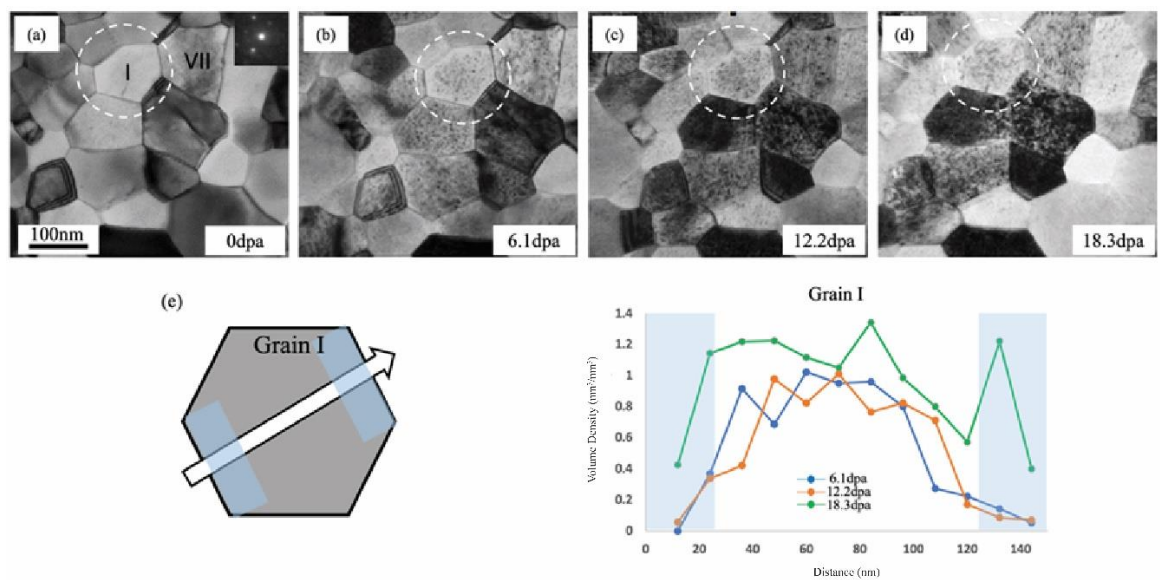

**Figure 1. Microstructure of the Irradiated Sample.** (a) Bright-field TEM image of a free-standing nanocrystalline pure Fe film prior to irradiation. (b-d) Bright-field TEM image of the Fe film irradiated up to 18.3 dpa. At high irradiation doses of 18.3 dpa, Defect denuded zones were found to collapse. (e) The defect volume density against distance away from the left bottom GB is plot. Blue line represents the sample irradiated at 6.1 dpa. Orange line represents the sample irradiated at 12 dpa. Green line represents the sample irradiated at 18.3 dpa.

Lending credibility to the notion that metastable forms of a GB result from irradiation is the striking observation that, at certain doses, some GBs exhibit denuded zone collapse, or shrinkage occurs with no observable changes in the GB macroscopic DOFs nor GB migration (**Figure. 2(a)**). While the measured macroscopic DOFs shown in Fig. S1 and Table S1 do not provide an explanation for the GB-dependent denuded zone collapse, analysis of the geometrically necessary dislocation (GND) in the GB region using Nye tensor mapping, described in the *Methods* section as well as in ref. [50], reveal a trend of increasing GND density with shrinking denuded zone (**Figure. 2(b)**). Clearly, then, to understand the time-dependent nature of the denuded zone near a GB, one needs to connect the internal structure of the GB, defined by a series of potentially metastable microstates/disconnection distribution with its ability to absorb damage.

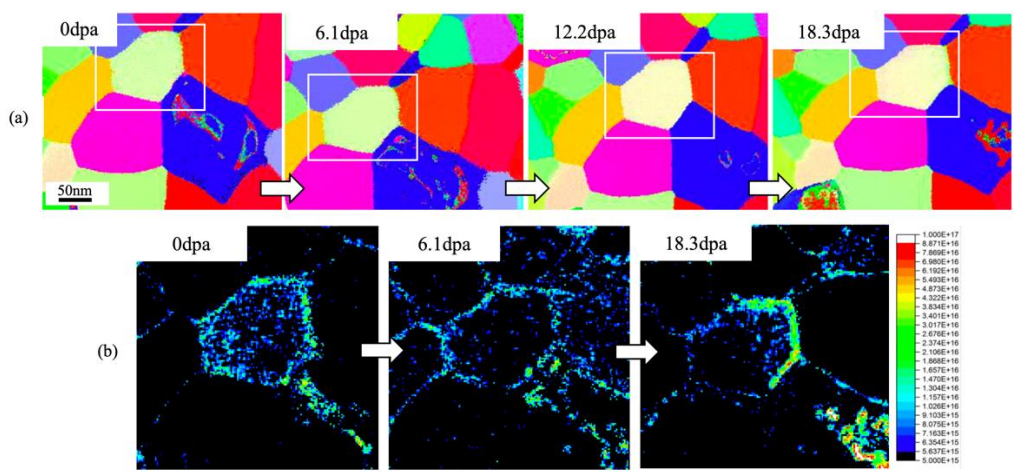

**Figure 2. Texture and Nye Tensor Mapping.** (a) ACOM maps of Fig. 2(a-d). (b) GND maps using calculation of Nye Tensor signals in the grain irradiated from 0 to 18.3 dpa. The variation of Nye tensor signals as increasing irradiation dose

Several models have been proposed to describe the mechanisms by which grain boundaries absorb vacancies and self-interstitial atoms (SIA) during irradiation. These include (1) enhanced vacancy/SIA recombination at the grain boundary, (2) climb of dislocations in the GB plane, and (3) point defect enhanced change of GB thermodynamic (micro)states. A 'micro-state' corresponds to a rigid translation of the two grains meeting at the GB. The first model is predicated on the preferential segregation of point defects to GBs and enhanced diffusion along the GB [19,51]. The second has been widely investigated for low-angle tilt GBs [52,53], and only recently for high-angle ones [54]. The third is associated with the recent realization that GBs usually exhibit a large number of metastable structures [12] and that point defect absorption can lead to transitions between these metastable states [11,19,23,25,55]. While each model implies that GBs change during irradiation/sink-operation, the difference in the underlying mechanisms has a profound effect on how the GB evolves under different conditions and the subsequent influence on the point defect field within a polycrystal. For example, the first two classes of models can only capture 'monotonic' GB property evolution, i.e., changes in the same direction of the capacity of GBs to absorb damage. We build on the flexibility afforded by model (3) to derive a model of GB evolution by interactions with irradiation defects. We note that this does not, necessarily, exclude the possibility that the other mechanisms may also operate, possibly modifying some details of the model presented here.

While bi-crystallography depends on the five macroscopic DOF that define a GB, these do not constrain GBs microscopic DOFs. **Figure 3** shows an example of a $\Sigma 5$ (310) [001] symmetric tilt GB in W at 1500 K before and after a large number of SIAs are placed into the GB and annealed. This necessarily involves displacing one crystal relative to the other in the direction normal to the GB to accommodate the extra volume incurred by defect absorption; this leads to an "elastic transfer" of much of this excess volume to the outside of the sample, as occurs in sintering large polycrystalline structures [55]. Recent atomistic simulations and statistical mechanics analyses have demonstrated that the number of metastable states increases with the reciprocal density of coincident sites $\Sigma$ of the GB [12]. $\Sigma$ is determined by the crystal structure and the misorientation of the grains (3 macroscopic DOFs). Another example of the atomistic changes in GB microstates brought about by the sustained absorption or emission of point defects is shown in Supplementary Fig. S2.

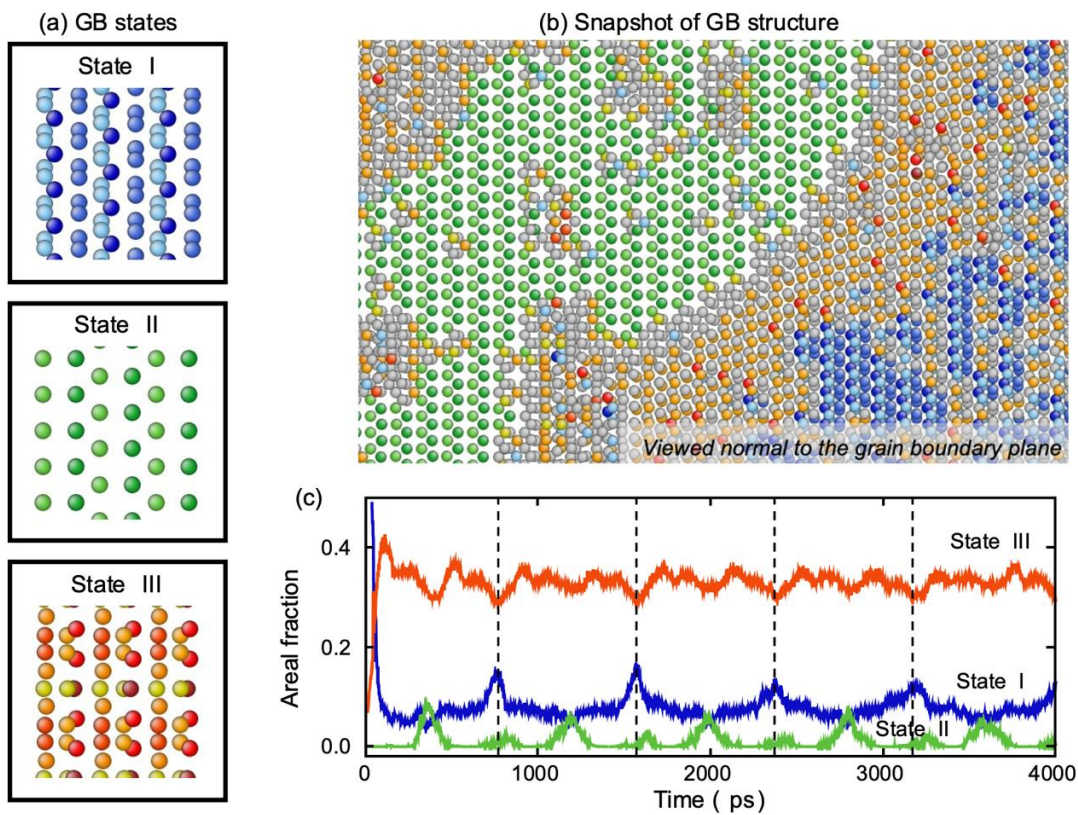

**Figure 3. Grain Boundary Structure and Microstates.** (a) Three stable/metastable states of a Σ5 (310) [001] symmetric tilt GB in W (labeled I, II, III). The colors label classes of topological type and are arbitrary. (b) Typical snapshot of GB structure during insertion of SIAs. Atoms are colored by same scheme as (a). (c) Time evolution of the areal fractions of the three states in (a). The dashed lines delimit the period. (This figure is reproduced from [55].)

Because all these processes are eminently atomistic in nature, next we provide evidence from molecular dynamics (MD) simulations of internal GB transitions mediated by point defect absorption/emission processes. As we will show, these transitions represent quantized energy jumps through different microstates that do not alter the global macro GB descriptors. The simulations are carried out by loading Σ5 and Σ29 boundaries with point defects at different rates to analyze the relaxation dynamics of these GBs as a function of the defect fluence. A detailed description of other important simulation parameters such as the interatomic potential used, the crystallographic structure of the simulation supercell, and the MD simulation conditions are given within the *Methods* section at the end of the paper.

To reflect a realistic scenario in irradiated metals, where an unbalanced flux of SIA and SIA clusters arrive to the GB, in the simulations we load the grain boundary with single-interstitials at different prescribed rates at a fixed temperature. **Figures 4a** and **4b** shows the number of defects absorbed vs. emitted as a function of exposure time (dose) at different rates in the Σ5 and Σ29 boundaries. **Figures 4c** and **4d** show the associated 'excess' GB energies (defined as the increase in GB energy with respect to the lowest energy microstate of the unirradiated GB). Considering the known shortcomings of MD simulations in terms of deposition rates (much faster than in any experiment), the simulations show that grain boundaries attempt to accommodate point defects by adopting metastable states compatible with the extra atomic density. Slower deposition rates allow more time for boundaries to structurally accommodate point defects and energetically minimize the new configuration into a metastable state. Conversely, increasing the rate of bombardment may force the grain boundary to reject point defects through emission events, due to insufficient time to find the next allowable metastable state before the next deposition event takes place. As Figures 4a and 4b illustrate, after a rapid energy increase upon the first few absorption events, the GB embarks on a dynamic relaxation process characterized by discrete energy jumps. Ostensibly, these jumps represent transitions between metastable states, which can be viewed as the minimum energy configuration of the combined ground state GB plus the corresponding number of absorbed defects. Generally, slower loading rates allow the boundary to accommodate higher concentrations of defects by assimilating them into allowable microstates characterized by higher GB energies. In contrast, fast defect arrival rates cannot be assimilated into the GB structure by internal relaxation processes, leading to point defect rejection and keeping the boundary closer to its original pristine structure.

The oscillations seen in the curves for the lower defect injection rates (125 and 250) in Figs. 4c and 4d are the signature of a spectrum of internal relaxation times, $\tau_R$, required by the GB to settle into a given microstate. This spectrum is GB and material dependent but effectively connects the internal relaxation dynamics to (which is governed by atomistic processes) to the evolution of the denuded zone. Effectively, the inverse of the defect arrival rate acts as a cutoff for longer structural relaxation times, preventing grain boundaries from accessing microstates that are only reachable through those longer transformations. Relaxation kinetics in GBs has been likened to glassy dynamics in disordered materials described by the Adam-Gibbs theory of structural glasses [12,23,57-60].

**Figure 4(e-f)** shows a comparison between the $\Sigma 29$ and $\Sigma 5$ grain boundaries for a fixed defect exposure rate and temperature. Based on this partial comparison, our results suggest that higher misorientation angles accelerate structural rearrangement, i.e., GB misorientation angles directly affect how point defects are accommodated under irradiation.

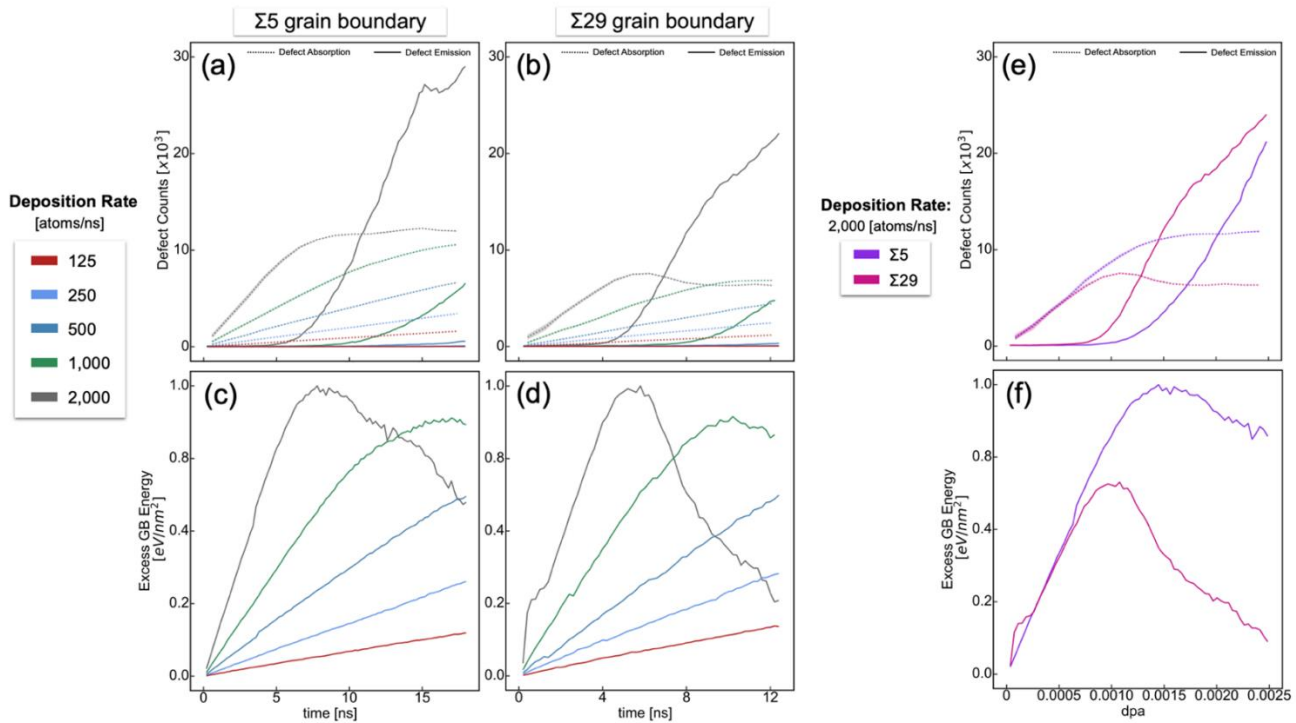

**Figure 4**. **Grain Boundary Response to Irradiation Defects.** (a-b) Defect counts as a function of time/dose, demonstrating the ability of a grain boundary to maintain constant absorption at lower rates of deposition. At high rates of deposition (1,000-2,000 atoms/ns), stagnated rates of absorption are exhibited, and the resultant emission of point defects is observed for both the (a) $\Sigma 5$ and (b) $\Sigma 29$ GBs. (c-d) Excess grain boundary energy as a function of time/dose, demonstrating the immediate spike in GB energy upon initial defect injection followed by the fluctuation of GB energy during continuous absorption for (a) $\Sigma 5$ and (b) $\Sigma 29$ GBs. Note the stark decrease in GB energy at the instant of GB absorption rate change and defect emission in (a-b). (e) $\Sigma 29$ grain boundaries experience defect emission before $\Sigma 5$ boundaries, suggesting an earlier saturation of microstates in $\Sigma 29$. (f) $\Sigma 5$ boundaries are able to reach much higher energies than $\Sigma 29$, indicating the existence of a wider range of allowable metastable energy states.

In summary, the MD simulations confirm that each GB (as defined by its macroscopic descriptors) displays a different hierarchy of microstates/relaxation times, and so different responses are to be expected from different boundaries.

**Discussion**

The results presented here provide a global picture for the interaction between GBs and irradiation defects under irradiation. In particular, point defect absorption modifies the structure of the boundary, that is, it affects GB microstates and thus likely the ability of the GB to absorb additional point defects. Without necessarily excluding other potential mechanisms, our work is general enough to explain the relationship between GB structure, irradiation conditions, material microstructure, and evolution of the denuded zone. In the following, we discuss the most salient features of our results and their connection to experimentally observable features of the microstructure.

*A unified picture of denuded zone onset and collapse*: Our analysis shows that the interplay between defects and GBs is governed by a relaxation time that reflects the instantaneous metastable structure of the grain boundary. In reality, this relaxation time is one of several time constants characterizing transitions between different microstates upon the absorption of point defects and defect clusters by the GB. Within this picture, when the characteristic times describing the irradiation process, e.g., the damage production rate and the GB absorption rate, are shorter than the corresponding GB relaxation times, the GB cannot adjust its structure (i.e., transition to a different metastable microstate) sufficiently fast to accommodate the damage internally. The implications of this picture as it relates to irradiation processes are:

(i) In general, each GB displays a specific $\tau_R$-spectrum representing the different relaxation times required to (meta) stabilize different microstates throughout its internal evolution with irradiation.

(ii) Zero microstates implies no defect absorption, i.e., zero sink strength, and thus the impossibility of denuded zone onset. An infinite number of microstates means that the GB behaves as a perfect absorber, which precludes the possibility of denuded zone collapse. Real boundaries fall between these two limits, dynamically shifting from one metastable state to another as dictated by transitions induced by irradiation.

(iii) When the irradiation process exhausts the short time region of the $\tau_R$-spectrum, the longer relaxation times reduce the admissible (absorbable) defect flux, gradually transitioning the system towards DZ collapse.

(iv) Because each GB type (defined by its macro-DOFs) displays its own set of characteristic relaxation times, they will respond differently to the same irradiation conditions. This explains why under identical irradiation conditions, e.g., particle type, dose rate, irradiation temperature, etc., denuded zone onset and collapse occur at different doses for different boundaries.

(v) Finally, an alternative scenario can be contemplated whereby –even if the GB can dynamically absorb defects– the resulting microstates may be too high in energy to be favored thermodynamically. This is further discussed below.

We find all of these points to be qualitatively consistent with our experimental observations.

*Connection to Nye-tensor measurements*: As discussed, once a given GB can no longer absorb arriving irradiation defects, the DZ begins to shrink until its eventual collapse. At that point, defects cannot be absorbed at a pace sufficiently fast to allow the GB to switch to other metastable microstates. This leads to two potentially distinct scenarios:

1. The first scenario directly comes from consideration of the different intrinsic relaxation times in the corresponding GB structure. Once the relaxation time is no longer short enough to evolve microstates before the next defect arrival time, defects will start to pile up against the GB and form clusters (when defect concentration is above the critical concentration for dislocation loop formation).

2. As advanced in point (v) above, the second scenario is one where the denuded zone can be interpreted as a thermodynamic structure where the free energy of the combined (GB + defect cluster) system is higher than the free energy of the GB in a different microstate that results from the prior microstate plus the absorption of the defect. Within this picture, the *next* absorption of an irradiation defect may result in a microstate that has a higher energy than the energies of the separate GB and defect structures. This crossover signals the transition towards the collapse of the DZ by accumulation of defects in the region adjacent to the GB.

The common consequence of these two scenarios is that the accumulation of defect clusters near the GB will inevitably result in reactions among these defects, giving rise to a stable 'network' of defects that contributes a measurable GND signal that can be observed experimentally. An example of the equivalence between dislocation segments and GND fingerprints is provided in Figure. 5 (see *Methods* for details). In this fashion, the existence of a Nye tensor signal is directly and intimately linked to fundamental GB properties and the kinetics of defect absorption by the GB in the context of the existing irradiation conditions. In other words, detection of a Nye-tensor signal next to a GB is unequivocally indicative of the collapse of the DZ.

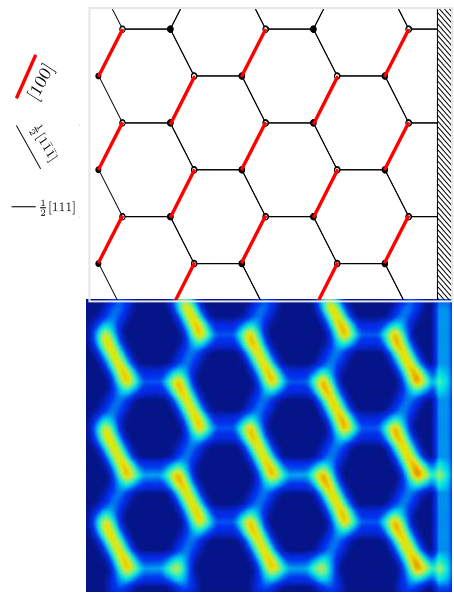

**Figure 5**. **GND Fingerprint of Defect Network.** Line (top) and equivalent GND scalar fingerprint (bottom) of a schematic dislocation network abutting a grain boundary (shaded band on the right).

Role of 'bulk' bias and other in-grain *irradiation processes*: The difference between the arrival rates of SIAs and vacancies to the GB may be associated with *bulk* bias, i.e., a differential feature in how SIAs and vacancies behave. The MD simulations have been carried out only with SIAs but no vacancies, which represents the ideal case of an infinite bias. However, in metallic materials the realistiuc bias emerges due to a number of different processes, whether working in isolation or in conjunction with one another, such as:

(i) The intrinsic defect diffusion dimensionality: SIAs and SIA clusters in BCC transition metals are known to move almost athermally (stages I and II in isochronal annealing) in one dimension. This allows them to rapidly move away from the cascade region (where they are produced in equal numbers to vacancies). By itself, this intrinsic bias is weak, and leads to small values of $c$.

(ii) Helium (or, more generally, 'noble gas atom') bias: This effect is caused by the presence of insoluble noble gas atoms (e.g., from direct ion implantation or neutron transmutation), which quickly segregate to vacancy-type defects, creating very stable immobile vacancy-He complexes that trap vacancies in the cascade region, limiting recombination and favoring SIA-defect escape from the cascade region towards sinks. This represents a very strong bias, and would naturally lead to a relatively high excess concentration $c$. In the *Supplementary Information* we provide additional simulations in support of this mechanism (Supplementary Fig. S3). Note that when the He concentration is too large to be accommodated into vacancy-type defects, He atoms can migrate directly to GBs, resulting in other types of modification of the GB structure/properties. However, the results from the kMC study described in the *Supplementary Information* section confirm self-interstitial atoms arrive at a GB in much larger proportions than vacancies, supporting the existence of a bulk bias favoring the arrival of SIAs to the boundary through simulated observation consequent to the He atom energetic properties within the Fe lattice. Experimentally, small He bubbles are witnessed within the microstructure at uniform distribution among the grains and grain boundaries, described in detail in El-Atwani et al. [7]. Bubble formation enhances loop formation through the process of loop-punching and trap-mutation, especially when helium vacancy complexes have limited mobility at the current experimental temperature.

(iii) Solute atom bias: substitutional solute atoms can be mobilized by coupling to point defect fluxes in what is known as *radiation enhanced diffusion* (RED). This coupling can be direct (point defects and solutes moving in the same direction) or inverse (opposite direction). This bias is behind *radiation enhanced segregation*; as non-equilibrium vacancy fluxes march towards sinks, depositing solute atoms in enhanced concentrations there. The coupling of vacancies and solutes can 'free' self-interstitials from recombination, contributing positively to the bias.

All of these potential biases, however, do not physically dictate, to first order, GB properties and how they contribute to sink efficiency.

*Using GB engineering to increase irradiation resistance*: Although the majority of the considerations are qualitative in nature, they provide both understanding and design guidelines for the development of materials with enhanced radiation tolerance. For example, a pre-characterization using atomistic models of the catalog of available microstates for a given set of (five-dimensional) macro DOFs can provide relative relaxation time spectra to gain an understanding what GBs are desirable and under what conditions to prevent DZ collapse. Conversely, one may design a thermomechanical process that favors the formation of microstructures with GBs known to bolster the GB sink efficiency. This can conceivably be achieved by directional solidification, rolling or extrusion of pre-formed internal microstructures (e.g., after achieving a target nanocrystalline structure).

Ultimately, while some microstructures containing GBs with line defect (disconnection) content structures can alter sink efficiency, to some extent; these are unlikely to be the controlling factors for establishing materials design protocols. The present study helps clarify the principles under which materials can be actively designed for radiation tolerance.

This work defines a relationship between GB sink efficiency limits and the local defect landscape created by the irradiation dose. The combined quantitative *in situ* electron microscopy and atomistic simulations reveals a picture in which GBs negotiate a metastable energy landscape under irradiation, transitioning through a set of quantized structural microstates characterized by a spectrum of internal relaxation times. When the defect arrival rate is slower that the intrinsic relaxation time, the GB acts as an efficient absorber, whereas in the opposite scenario it behaves as if having zero sink efficiency. Our atomistic simulations, consistent with our experimental observations, conclusively show that when the defect flux overwhelms the internal relaxation dynamics, the GB acts as a saturated absorber, rejecting defects back into the bulk and signaling the collapse of the denuded zone.

The proposed picture based on GB microstate changes controlling further defect absorption, extends our understanding of the principles under which materials can be actively designed with radiation tolerance as a design criterion.

## Materials and Methods

### *Film Growth, Irradiation, and Characterization:*

Polycrystalline iron films with a target thickness of 100 nm were deposited using magnetron sputtering onto sodium chloride substrates to create nanocrystalline iron specimens suitable for transmission electron microscopy (see Ref. [61] for details). The samples were irradiated using 10 keV He$^+$ at Sandia National Laboratories via a 10 kV Colutron source. The final He/Fe atomic ratio of 0.41 within the implanted region as reported in El-Atwani et al. [27] based on SRIM simulations given a 40 eV displacement damage energy threshold [62]. Specifically, *in situ* irradiation was performed on a well-characterized, columnar, polycrystalline Fe film [61]. GBs were monitored at the microscale level (~ 100 nm) throughout irradiation using a combination of *in situ* bright field imaging by a highly modified JEOL 2100 LaB$_6$ TEM videos were captured at 200kV and a frame rate of 30 frames per second. Precession-electron-diffraction-enhanced automated crystallographic orientation mapping (ACOM) was taken intermittently at each dose step [7]. In order to track GB structure, sequences of TEM images were taken at an unchanging two-beam diffraction condition (see supplementary Fig. S4.) during *in situ* irradiation, allowing that the defect density at different irradiation stages can be estimated under the same imaging condition. The orientation data acquired by ACOM-TEM systems [63,64] was used as the input for GND density calculations [50]. GB planes were determined by observing the projected inclined GB trace in TEM coupled with the ACOM data [65]. Denuded zones were determined from the bright field TEM images/videos, visualized in the supplementary Fig. S5 following methodologies outlined in [7,9,27]. ACOM [63,64] was performed in steps throughout He$^+$ implantation up to ~18 dpa to deduce the five macro-DOFs of the GBs [66].

### *Molecular Dynamics Simulations:*

We have used the Large-Scale Atomic/Molecular Massively Parallel Simulator (LAMMPS) [67] to construct two pure iron grain boundaries of different misorientation angle. These angles were 36.9° and 43.6°, corresponding to Σ5 and Σ29 grain boundaries, respectively. The computational supercell was constructed as a bcc bi-crystal using a lattice constant of a = 2.8665 Å and cell dimensions of about 28 nm x 38 nm x 38 nm, resulting in approximately 2.5 million atoms. The grain boundary area in this system made up around 4% of the total supercell (34,500 atoms). After both systems were equilibrated at 550K, self-interstitial atoms were deposited into the boundaries to observe the effect of point defect loading rate on boundary accommodation. Five independent boundary loading simulations were performed for each grain boundary at a finite temperature of 550K, with the rate of deposition changing between each simulation. The rates were 125, 250, 500, 1,000, and 2,000 [atoms/ns]. Grain boundary emission and absorption counts were tracked as the depositions reached proper timescales, demonstrating the stability of defect absorption until a boundary approaches an overloaded state. The potential energy of each boundary was computed throughout each deposition to observe a lowering of energy with grain boundary accommodation of the newly introduced point defects.

In all MD simulations, we used the interatomic potential for bcc Fe developed by Dudarev and Derlet [68], which is formulated as an embedded-atom method (EAM) that captures magnetic effects while retaining its functionality for large scale MD simulations. The Dudarev-Derlet potential was specifically developed for irradiation damage simulations and is fitted to DFT calculations of point defect energies for both the ferromagnetic and non-magnetic phases of $\alpha$-Fe at zero temperature.

### *Atomistic Parameterization of the Kinetic Monte Carlo Simulations*

We also have used atomistic simulations to provide insight into how changing microstates impact the interaction of defects with grain boundaries, essentially to address the assumptions made in the kinetic Monte Carlo (kMC) modeling. Specifically, two effects were probed for one specific grain boundary (Σ13 [111](134) symmetric tilt). First, the highest and lowest energy structures found by mapping the gamma surface were examined, representing different translational relationships between the two grains. Second, the lowest energy structure was modified ("loaded") by adding 10 SIAs, one at a time, with 2.0 picosecond anneals at 750 K and minimizations between each insertion. In all three resulting structures, the segregation energy of vacancies was calculated by removing atoms, one-by-one, and minimizing the energy of the structure.

### *Kinetic Monte Carlo Simulations:*

The results shown in Fig. S3 pertain to an object kMC model in which single defects are treated explicitly. The grain boundary is modeled as a region in which the defect properties are modified. While in principle both the thermodynamic and kinetic properties of defects can vary at the boundary plane, in the kMC calculations only the binding energy was changed, meaning that defects had an attractive binding energy to the boundary plane as a higher activation energy for the defect to exit the boundary compared to the barrier to enter, implying boundaries are saturable as there is a finite probability for defects to exit. The kMC model was used to study two scenarios, one in which the only mechanisms for defect removal was interstitial-vacancy recombination, whether in the bulk or the boundary, and a second in which there was a bulk bias for vacancies (vacancies in the bulk were annihilated at an enhanced rate compared to interstitials). All the parameters in the kMC model are given in [69,70]. Most notably, self-interstitials had a binding energy of 0.8 eV to the GB, while vacancies were characterized by no attraction to the boundary. Vacancies were eliminated at a rate of $k^2 D_v N_v$, with $k^2 = 10^{12}$ cm$^{-2}$, $N_v$ the number of vacancies in the system, mimicking the He$_{int}$ + Vacancy → He$_{sub}$ reaction. We used a dose rate of $10^{-2}$ dpa/s for the generation of Frenkel pairs and a temperature of 500 K.

### *Grain Boundary Structure and Nye Tensor Analysis:*

One of the principal signatures of denuded zone collapse is the emergence of a nonzero GND signal. Experimentally, GNDs are detected by mapping the gradient in lattice curvature to Nye's definition of the GND tensor [50,71,72]: $\boldsymbol{\alpha} = \boldsymbol{\kappa}^T - \text{Tr}(\boldsymbol{\kappa})\boldsymbol{I}$, where $\boldsymbol{\kappa} = \partial \boldsymbol{\omega}/\partial x$ is the lattice curvature tensor and $\boldsymbol{\omega}$ is the local orientation matrix. The challenge when interpreting these signals emerges from reconciling crystal orientation measurements with the dislocation-based definition of $\boldsymbol{\alpha}$, i.e., $\boldsymbol{\alpha} = \sum_m \boldsymbol{b}_m \otimes \boldsymbol{t}_m$, where $\boldsymbol{b}$ and $\boldsymbol{t}$ are the Burgers vector and line direction of a dislocation segment $m$, which runs over the entire set of discrete dislocation segments representing a given dislocation substructure. This formula represents an unambiguous definition of the GND, as it represents the net Burgers vector of specific dislocation configurations whose structure is known. Accordingly, a nonzero signal must necessarily correspond to a set of unbalanced loops, i.e., loops that have partially lost part of their perimeter owing to some absorption and/or reaction mechanism. While this is reasonable in the vicinity of saturated or near-saturated grain boundaries, our calculations show unequivocally that the discretization of space used to compute $\boldsymbol{\alpha}$ in either of its two versions (curvature-based in experiments or dislocation-segment based in the simulations) also can drastically alter the final GND signal. This introduces an extra dimensional complexity that must be accounted for. In other words, full prismatic loops that overall contribute no GND signature can indeed yield nonzero values for sufficiently fine pixelations of the microstructure. As such, several scenarios must be contemplated when comparing both definitions of $\boldsymbol{\alpha}$ on a discrete grid to assess the practical loop/grain boundary configurations that may give rise to a nonzero signal:

(i) First, we look at atomistic configurations of bulk, partially absorbed, and fully-absorbed loops, and obtain their corresponding Nye tensor signals. For consistency with the structure of irradiation defect clusters in bcc metals, we consider 1/2 ⟨111⟩ loops arriving at a virgin grain boundary (full defect absorbance) via diffusion from the grain's interior. Our main observation is that indeed partially absorbed loops do yield a nonzero GND signal, regardless of the discretization resolution [73]. This suggests that nonzero GND footprints must be due to loop structures that have been partially absorbed or that have suffered Burger's vector reactions such that an unbalanced total Burgers vector is produced.

(ii) Second, we analyzed full closed prismatic loops with various degrees of resolution. The conclusion is that nonzero GND signals can indeed be produced for sufficiently fine discretizations. This requires a careful evaluation of the expected or measured loop size vis-à-vis the microscopy pixelation resolution, to ensure that nonzero signals are properly interpreted. This also suggests that a scalar metric representing the Nye tensor (such as the L1 norm) may not be suitable for identifying GNDs [74–77].

We employ models to distinguish between these two possibilities, to construe Nye tensor information and to analyze Nye tensor data [50] in order to understand the connection between GB defects and the denuded zone.

# References

1. G.J. Wang, A.P. Sutton, V. Vitek, A computer simulation study of <001> and <111> tilt boundaries: the multiplicity of structures, Acta Metallurgica. 32 (1984) 1093–1104.
2. G.A. Vetterick, J. Gruber, P.K. Suri, J.K. Baldwin, M.A. Kirk, P. Baldo, Y.Q. Wang, A. Misra, G.J. Tucker, M.L. Taheri, Achieving Radiation Tolerance through Non-Equilibrium Grain Boundary Structures, Scientific Reports. 7 (2017) 12275.
3. C.M. Barr, G.A. Vetterick, K.A. Unocic, K. Hattar, X.-M. Bai, M.L. Taheri, Anisotropic radiation-induced segregation in 316L austenitic stainless steel with grain boundary character, Acta Materialia. 67 (2014) 145–155.
4. M. Samaras, P.M. Derlet, H. Van Swygenhoven, M. Victoria, Atomic scale modelling of the primary damage state of irradiated fcc and bcc nanocrystalline metals, Journal of Nuclear Materials. 351 (2006) 47–55.
5. J. Wang, R.G. Hoagland, J.P. Hirth, A. Misra, Atomistic modeling of the interaction of glide dislocations with "weak" interfaces, Acta Materialia. 56 (2008) 5685–5693.
6. S.J. Zinkle, L.L. Snead, Designing Radiation Resistance in Materials for Fusion Energy, Annual Review of Materials Research. 44 (2014) 241–267.
7. O. El-Atwani, J.E. Nathaniel, A.C. Leff, K. Hattar, M.L. Taheri, Direct Observation of Sink-Dependent Defect Evolution in Nanocrystalline Iron under Irradiation, Scientific Reports. 7 (2017) 1836.
8. W.Z. Han, M.J. Demkowicz, E.G. Fu, Y.Q. Wang, A. Misra, Effect of grain boundary character on sink efficiency, Acta Materialia. 60 (2012) 6341–6351.
9. O. El-Atwani, J.E. Nathaniel, A.C. Leff, J.K. Baldwin, K. Hattar, M.L. Taheri, Evidence of a temperature transition for denuded zone formation in nanocrystalline Fe under He irradiation, Materials Research Letters. 5 (2017) 195–200.
10. A. Das, Grain boundary engineering: fatigue fracture, Philosophical Magazine. 97 (2017) 867–916.
11. T. Frolov, W. Setyawan, R.J. Kurtz, J. Marian, A.R. Oganov, R.E. Rudd, Q. Zhu, Grain boundary phases in bcc metals, Nanoscale. 10 (2018) 8253–8268.
12. J. Han, V. Vitek, D.J. Srolovitz, Grain-boundary metastability and its statistical properties, Acta Materialia. 104 (2016) 259–273.
13. A. Dunn, R. Dingreville, E. Martínez, L. Capolungo, Identification of dominant damage accumulation processes at grain boundaries during irradiation in nanocrystalline α-Fe: A statistical study, Acta Materialia. 110 (2016) 306–323.
14. R. Sutton, A.P., Ballufii, Interfaces in Crystalline Materials. Oxford, Clarendon Press; Oxford University Press, New York, 1995.
15. M. Samaras, W. Hoffelner, M. Victoria, Irradiation of pre-existing voids in nanocrystalline iron, Journal of Nuclear Materials. 352 (2006) 50–56.
16. R.E. Stoller, P.J. Kamenski, Yu.N. Osetsky, Length-scale Effects in Cascade Damage Production in Iron, MRS Proceedings. 1125 (2008) 1105–1125.
17. R. Yamada, S.J. Zinkle, G. Philip Pells, Microstructure of $Al_2O_3$ and $MgAl_2O_4$ preimplanted with H, He, C and irradiated with $Ar^+$ ions, Journal of Nuclear Materials. 209 (1994) 191–203.
18. S.J. Zinkle, Microstructure of ion irradiated ceramic insulators, Nuclear Instruments and Methods in Physics Research Section B: Beam Interactions with Materials and Atoms. 91 (1994) 234–246.
19. Q. Zhu, A. Samanta, B. Li, R.E. Rudd, T. Frolov, Predicting phase behavior of grain boundaries with evolutionary search and machine learning, Nature Communications. 9 (2018) 467.

20. X. Zhang, K. Hattar, Y. Chen, L. Shao, J. Li, C. Sun, K. Yu, N. Li, M.L. Taheri, H. Wang, J. Wang, M. Nastasi, Radiation damage in nanostructured materials, Progress in Materials Science. 96 (2018) 217–321.
21. G. Meric de Bellefon, I.M. Robertson, T.R. Allen, J.-C. van Duysen, K. Sridharan, Radiation-resistant nanotwinned austenitic stainless steel, Scripta Materialia. 159 (2019) 123–127.
22. Y. Oh, V. Vitek, Structural multiplicity of $\sum = 5(001)$ twist boundaries and interpretation of X-ray diffraction from these boundaries, Acta Metallurgica. 34 (1986) 1941–1953.
23. T. Frolov, D.L. Olmsted, M. Asta, Y. Mishin, Structural phase transformations in metallic grain boundaries, Nature Communications. 4 (2013) 1899.
24. J. Han, V. Vitek, D.J. Srolovitz, The grain-boundary structural unit model redux, Acta Materialia. 133 (2017) 186–199.
25. J. Han, V. Vitek, D.J. Srolovitz, The interplay between grain boundary structure and defect sink/annealing behavior, IOP Conference Series: Materials Science and Engineering. 89 (2015) 12004.
26. B.P. Uberuaga, L.J. Vernon, E. Martinez, A.F. Voter, The relationship between grain boundary structure, defect mobility and grain boundary sink efficiency, Scientific Reports. 5 (2015) 9095.
27. O. El-Atwani, J.E. Nathaniel, A.C. Leff, B.R. Muntifering, J.K. Baldwin, K. Hattar, M.L. Taheri, The role of grain size in He bubble formation: Implications for swelling resistance, Journal of Nuclear Materials. 484 (2017) 236–244.
28. C.M. Barr, S. Thomas, J.L. Hart, W. Harlow, E. Anber, M.L. Taheri, Tracking the evolution of intergranular corrosion through twin-related domains in grain boundary networks, Npj Materials Degradation. 2 (2018) 14.
29. K. Beierschmitt, M. Buchanan, A. Clark, I. Robertson, P. Britt, A. Navrotsky, P. Burns, P. Tortorelli, A. Misra, J. Wishart, P. Fenter, A. Gewirth, B. Wirth, B. Mincher, I. Szlufarska, J. Busby, L. Horton, B. Garrett, J. Vetrano, P. Wilk, K. Runkles, S. Kung, S. Lesica, B. Wyatt, D. Counce, K. Jones, Basic Research Needs for Future Nuclear Energy: Report of the Basic Energy Sciences Workshop for Future Nuclear Energy, August 9-11, 2017, 2017.
30. Z. Bai, G.H. Balbus, D.S. Gianola, Y. Fan, Mapping the kinetic evolution of metastable grain boundaries under non-equilibrium processing, Acta Materialia. 200 (2020) 328–337.
31. C.M. Barr, O. El-Atwani, D. Kaoumi, K. Hattar, Interplay Between Grain Boundaries and Radiation Damage, JOM. 71 (2019) 1233–1244.
32. J. Wei, B. Feng, R. Ishikawa, T. Yokoi, K. Matsunaga, N. Shibata, Y. Ikuhara, Direct imaging of atomistic grain boundary migration, Nature Materials. 20 (2021) 951–955.
33. C. Sun, M. Song, K.Y. Yu, Y. Chen, M. Kirk, M. Li, H. Wang, X. Zhang, In situ Evidence of Defect Cluster Absorption by Grain Boundaries in Kr Ion Irradiated Nanocrystalline Ni, Metallurgical and Materials Transactions A. 44 (2013) 1966–1974.
34. P. Yan, J. Zheng, J. Liu, B. Wang, X. Cheng, Y. Zhang, X. Sun, C. Wang, J.-G. Zhang, Tailoring grain boundary structures and chemistry of Ni-rich layered cathodes for enhanced cycle stability of lithium-ion batteries, Nature Energy. 3 (2018) 600–605.
35. P.-J. Chang, K.-Y. Cheng, S.-W. Chou, J.-J. Shyue, Y.-Y. Yang, C.-Y. Hung, C.-Y. Lin, H.-L. Chen, H.-L. Chou, P.-T. Chou, Tri-iodide Reduction Activity of Shape- and Composition-Controlled PtFe Nanostructures as Counter Electrodes in Dye-Sensitized Solar Cells, Chemistry of Materials. 28 (2016) 2110–2119.
36. K.S. Il, L.K. Hyoung, M.H. A, K.H. Sik, H.S. Woo, R.J. Wook, Y.D. Jin, S.W. Ho, L.X. Shu, L.Y. Hee, S.G. Jeffrey, K.S. Wng, Dense dislocation arrays embedded in grain boundaries for high-performance bulk thermoelectrics, Science. 348 (2015) 109–114.
37. I.J. Beyerlein, M.J. Demkowicz, A. Misra, B.P. Uberuaga, Defect-interface interactions, Progress in Materials Science. 74 (2015) 125–210.

38. M. Nastasi, N. Michael, J. Mayer, J.K. Hirvonen, M. James, Ion-solid interactions: fundamentals and applications, Cambridge University Press, 1996.
39. O. El-Atwani, E. Martinez, E. Esquivel, M. Efe, C. Taylor, Y.Q. Wang, B.P. Uberuaga, S.A. Maloy, Does sink efficiency unequivocally characterize how grain boundaries impact radiation damage?, Physical Review Materials. 2 (2018) 113604.
40. H. Gleiter, Grain boundaries as point defect sources or sinks—Diffusional creep, Acta Metallurgica. 27 (1979) 187–192.
41. M. Rose, A.G. Balogh, H. Hahn, Instability of irradiation induced defects in nanostructured materials, Nuclear Instruments and Methods in Physics Research Section B: Beam Interactions with Materials and Atoms. 127–128 (1997) 119–122.
42. C.M. Barr, L. Barnard, J.E. Nathaniel, K. Hattar, K.A. Unocic, I. Szlurfarska, D. Morgan, M.L. Taheri, Grain boundary character dependence of radiation-induced segregation in a model Ni–Cr alloy, Journal of Materials Research. 30 (2015) 1290–1299.
43. J. Li, K.Y. Yu, Y. Chen, M. Song, H. Wang, M.A. Kirk, M. Li, X. Zhang, In situ study of defect migration kinetics and self-healing of twin boundaries in heavy ion irradiated nanotwinned metals, Nano Letters. 15 (2015) 2922–2927.
44. M.J. Demkowicz, R.G. Hoagland, B.P. Uberuaga, A. Misra, Influence of interface sink strength on the reduction of radiation-induced defect concentrations and fluxes in materials with large interface area per unit volume, Physical Review B. 84 (2011) 104102.
45. R.W. Siegel, S.M. Chang, R.W. Balluffi, Vacancy loss at grain boundaries in quenched polycrystalline gold, Acta Metallurgica. 28 (1980) 249–257.
46. W. Krakow, Structural multiplicity observed at a Σ = 5/[001] 53·1° tilt boundary in gold, Philosophical Magazine A. 63 (1991) 233–240.
47. G. Henkelman, B.P. Uberuaga, H. Jónsson, A climbing image nudged elastic band method for finding saddle points and minimum energy paths, The Journal of Chemical Physics. 113 (2000) 9901–9904.
48. A.P. Sutton, Models of the atomic and electronic structures of grain boundaries in silicon, Structure and Properties of Dislocations in Semiconductors 1989. (1989) 13.
49. Y. Yang, C.A. Dickerson, H. Swoboda, B. Miller, T.R. Allen, Microstructure and mechanical properties of proton irradiated zirconium carbide, Journal of Nuclear Materials. 378 (2008) 341–348.
50. A.C. Leff, C.R. Weinberger, M.L. Taheri, Estimation of dislocation density from precession electron diffraction data using the Nye tensor, Ultramicroscopy. 153 (2015) 9–21.
51. X.-M. Bai, A.F. Voter, R.G. Hoagland, M. Nastasi, B.P. Uberuaga, Efficient Annealing of Radiation Damage Near Grain Boundaries via Interstitial Emission, Science. 327 (2010) 1631–1634.
52. Y. Gu, J. Han, S. Dai, Y. Zhu, Y. Xiang, D.J. Srolovitz, Point defect sink efficiency of low-angle tilt grain boundaries, Journal of the Mechanics and Physics of Solids. 101 (2017) 166–179.
53. Y. Zhu, J. Luo, X. Guo, Y. Xiang, S.J. Chapman, Role of Grain Boundaries under Long-Time Radiation, Physical Review Letters. 120 (2018) 222501.
54. S. Chu, P. Liu, Y. Zhang, X. Wang, S. Song, T. Zhu, Z. Zhang, X. Han, B. Sun, M. Chen, In Situ Atomic-Scale Observation of Dislocation Climb and Grain Boundary Transformation in Nanostructured Metal, Research Square. PREPRINT ( (2021).
55. E.A. Lazar, J. Han, D.J. Srolovitz, Topological framework for local structure analysis in condensed matter, Proceedings of the National Academy of Sciences of the United States of America. 112 (2015) E5769–E5776.
56. F. Thümmler, W. Thomma, The sintering process, Metallurgical Reviews. 12 (1967) 69–108.

57. H. Zhang, D.J. Srolovitz, J.F. Douglas, J.A. Warren, Grain boundaries exhibit the dynamics of glass-forming liquids, Proceedings of the National Academy of Sciences. 106 (2009) 7735–7740.
58. S.R. Phillpot, D. Wolf, H. Gleiter, A structural model for grain boundaries in nanocrystalline materials, Scripta Metallurgica et Materialia. 33 (1995) 1245–1251.
59. P. Keblinski, D. Wolf, S.R. Phillpot, H. Gleiter, Structure of grain boundaries in nanocrystalline palladium by molecular dynamics simulation, Scripta Materialia. 41 (1999) 631–636.
60. T. Brink, K. Albe, From metallic glasses to nanocrystals: Molecular dynamics simulations on the crossover from glass-like to grain-boundary-mediated deformation behaviour, Acta Materialia. 156 (2018) 205–214.
61. G. Vetterick, J.K. Baldwin, A. Misra, M.L. Taheri, Texture evolution in nanocrystalline iron films deposited using biased magnetron sputtering, Journal of Applied Physics. 116 (2014).
62. J.-P. Crocombette, T. Jourdan, Cell Molecular Dynamics for Cascades (CMDC): A new tool for cascade simulation, Nuclear Instruments and Methods in Physics Research Section B: Beam Interactions with Materials and Atoms. 352 (2015) 9–13.
63. J. Portillo, E.F. Rauch, S. Nicolopoulos, M. Gemmi, D. Bultreys, Precession Electron Diffraction Assisted Orientation Mapping in the Transmission Electron Microscope, Materials Science Forum. 644 (2010) 1–7.
64. E.F. Rauch, J. Portillo, S. Nicolopoulos, D. Bultreys, S. Rouvimov, P. Moeck, Automated nanocrystal orientation and phase mapping in the transmission electron microscope on the basis of precession electron diffraction, Zeitschrift Für Kristallographie. 225 (2010) 103–109.
65. M.I. Hartshorne, A.C. Leff, G.A. Vetterick, E.M. Hopkins, M.L. Taheri, Grain Boundary Plane Measurement in Atom Probe Tomography and Transmission Electron Diffraction Specimens Using Precession Electron Diffraction, Microscopy and Microanalysis. (2022) Submitted.
66. J.E. Nathaniel, A.C. Lang, O. El-Atwani, P.K. Suri, J.K. Baldwin, M.A. Kirk, Y. Wang, M.L. Taheri, Toward high-throughput defect density quantification: A comparison of techniques for irradiated samples, Ultramicroscopy. 206 (2019) 112820.
67. LAMMPS - a flexible simulation tool for particle-based materials modeling at the atomic, meso, and continuum scales, A. P. Thompson, H. M. Aktulga, R. Berger, D. S. Bolintineanu, W. M. Brown, P. S. Crozier, P. J. in 't Veld, A. Kohlmeyer, S. G. Moore, T. D. Nguyen, R. Shan, M. J. Stevens, J. Tranchida, C. Trott, S. J. Plimpton, Comp Phys Comm, 271 (2022) 10817.
68. Chiesa S, Derlet PM, Dudarev SL, Swygenhoven HV. Optimization of the magnetic potential
for α-fe. Journal of physics. Condensed matter. 2011;23(20):206001-14.
69. O. Senninger, F. Soisson, E. Martínez, M. Nastar, C.-C. Fu, Y. Bréchet, Modeling radiation induced segregation in iron–chromium alloys, Acta Materialia. 103 (2016) 1–11.
70. E. Martínez, O. Senninger, C.-C. Fu, F. Soisson, Decomposition kinetics of Fe-Cr solid solutions during thermal aging, Physical Review B. 86 (2012) 224109.
71. R. Matsutani, S. Onaka, Representation of Nye's Lattice Curvature Tensor by Log Angles, MATERIALS TRANSACTIONS. 60 (2019) 935–938.
72. C. Begau, J. Hua, A. Hartmaier, A novel approach to study dislocation density tensors and lattice rotation patterns in atomistic simulations, Journal of the Mechanics and Physics of Solids. 60 (2012) 711–722.
73. T.J. Ruggles, D.T. Fullwood, J.W. Kysar, Resolving geometrically necessary dislocation density onto individual dislocation types using EBSD-based continuum dislocation microscopy, International Journal of Plasticity. 76 (2016) 231–243.

74. L.M. Woryk, S. He, E.M. Hopkins, C.-Y. Hung, J. Han, D.J. Srolovitz, J. Marian, M.L. Taheri, Geometrically necessary dislocation fingerprints of dislocation loop absorption at grain boundaries, Physical Review Materials. 6 (2022) 83804.
75. S. Das, F. Hofmann, E. Tarleton, Consistent determination of geometrically necessary dislocation density from simulations and experiments, International Journal of Plasticity. 109 (2018) 18–42.
76. A. Nye, A.C. Leff, C.M. Barr, M.L. Taheri, Direct observation of recrystallization mechanisms during annealing of Cu in low and high strain conditions, Scripta Materialia. 146 (2018) 308–311.
77. A.C. Leff, M.L. Taheri, Quantitative assessment of the driving force for twin formation utilizing Nye tensor dislocation density mapping, Scripta Materialia. 121 (2016) 14–17.
78. Yan-Ru Lin, Arunodaya Bhattacharya, Steven J. Zinkle, Analysis of position-dependent cavity parameters in irradiated metals to obtain insight on fundamental defect migration phenomena, Materials & Design, Vol. 226 (2023) 111668.
79. Yu.V. Konobeev, A.V. Subbotin, V.N. Bykov, V.I. Tscherbak, Grain boundary void denuded zone in irradiated metals, Physica Status Solidi (a) 29 K121 (1975).
80. Zinkle SJ, Farrell K. Void swelling and defect cluster formation in reactor-irradiated copper. *J Nucl Mater*. 1989;168(3):262-267.
81. Wadsworth, J., Ruano, O.A. & Sherby, O.D. Denuded zones, diffusional creep, and grain boundary sliding. Metall Mater Trans A 33, 219–229 (2002).

**Acknowledgments**

The authors acknowledge funding in part from the US Department of Energy, Office of Basic Energy Sciences through contract DE-SC0008274 (M.L.T., O.E, A.C.L, J.E.N.) and contract DE-SC0020314 (M.L.T., C-Y.H., E.M.H., J.M., S.H. L.W.). Initial work by DS was supported by BES with follow-on research supported by Hong Kong Research Grants Council Collaborative Research Fund C1005-19G. JH acknowledges support from the Early Career Scheme (ECS) of Hong Kong RGC Grant 9048213 and Donation for Research Projects 9229061. K.H. was supported by the DOE-BES Materials Science and Engineering Division under FWP 15013170. This work was performed, in part, at the Center for Integrated Nanotechnologies, an Office of Science User Facility operated for the U.S. Department of Energy (DOE) Office of Science. Sandia National Laboratories is a multi-mission laboratory managed and operated by National Technology and Engineering Solutions of Sandia, LLC, a wholly owned subsidiary of Honeywell International, Inc., for the U.S. DOE's National Nuclear Security Administration under contract DE-NA-0003525. The views expressed in the article do not necessarily represent the views of the U.S. DOE or the United States Government. The authors acknowledge J. Kevin Baldwin of the Center for Integrated Nanotechnologies at Los Alamos National Laboratory for film growth for this study. B.P.U., who performed the atomistic calculations, acknowledges support by FUTURE (Fundamental Understanding of Transport Under Reactor Extremes), an Energy Frontier Research Center funded by the U.S. Department of Energy (DOE), Office of Science, Basic Energy Sciences (BES). Los Alamos National Laboratory is operated by Triad National Security, LLC, for the National Nuclear Security Administration of U.S. Department of Energy (Contract No. 89233218CNA000001)

**Funding:**

US Department of Energy, Office of Basic Energy Sciences DE-SC0008274, DE-SC0020314

Hong Kong Research Grants Council Collaborative Research Fund C1005-19G
Early Career Scheme (ECS) of Hong Kong RGC Grant 9048213
Donation for Research Projects 9229061
DOE-BES Materials Science and Engineering Division FWP 15013170
U.S. DOE's National Nuclear Security Administration DE-NA-0003525
National Nuclear Security Administration of U.S. Department of Energy 89233218CNA000001

**Author contributions:**
    Conceptualization: MT, DS, JM, OE
    Methodology: OE, CYH, EH, AB, AL, JN, LW, JH, EM
    Investigation: MT, JM, OE, JH, BU, EM, DS
    Visualization: AB, OE, JN
    Supervision: MT, JM, BU, KH, DS
    Writing—original draft: OE, JM, JH, BU, EM, DS
    Writing—review & editing: MT, JM, AB, OE, BU, KH

**Competing interests:**
The authors declare that they have no competing interests.

**Data and materials availability:**
All data needed to evaluate the conclusions in the paper are present in the paper and/or the Supplementary Materials. Additional data related to this paper may be requested from the authors.

# Supplementary Materials for

## Grain boundary metastability controls irradiation resistance in nanocrystalline metals

Osman El Atwani *et al.*

*Corresponding author. Mtaheri4@jhu.edu

**This PDF file includes:**

Supplementary Text
Figs. S1 to S5
Tables S1
References (1 to 8)

**Supplementary Text**

**Orientation/GB character and Nye tensor analysis**

Automated Crystallographic Orientation Mapping (ACOM) was performed via NanoMEGAS ASTAR precession diffraction [4]. A spot size of 5 nm and a step size of 2 nm were used. The experimental electron diffraction spot patterns were acquired and compared to one or several sets of thousands of computer-generated electron diffraction spot patterns to obtain information of nanocrystal orientation/GB character at varying doses of 0, 6.1, 12.2, and 18.3dpa in the TEM [5]. The software technique for the comparisons is based on optimal template matching using cross-correlation. ACOM map was used to identify grains with compatible diffraction condition for dislocation loop imaging and to ensure diffraction conditions to be constant during irradiation.

The Nye tensor defined in Eq. 1 is described in terms of the contortion K, as follows [6]:

$$\alpha_{ij} = K_{ji} - \delta_{ij} K_{kk} \qquad (1)$$

where $K_{kk}$ is in the Einstein summation notation implying that $K_{kk} = \sum_i K_{ij} \cdot \delta_{ij} = 1$ for $i = j$ and $\delta_{ij} = 0$ for $i \neq j$. Nye defined the contortion as a second-rank tensor

$$K_{ij} = \frac{\partial \varphi_i}{\partial x_i} = \frac{\Delta \varphi_i}{\Delta x_i} \qquad (2)$$

, where $\partial \varphi_i$ is the three-dimensional rotation vector and $\partial x_i$ is the displacement vector. These are generated using ACOM-TEM. First, the axis-angle pair for the disorientation between each point and its neighbors is calculated. The rotation vector $\partial \varphi_i$ is approximated as the product of the axis and angle according to infinitesimal rotation theory [7]. $\partial x_i$ is the displacement between the steps being compared. Eq. 2 is undefined whenever $j = 3$ because the displacement along the z-axis is zero. There are undefined values for $K_{i3}$ will carry through when K is plugged into Eq. 1 [8], which is written explicitly as

$$\alpha_{ij} = \begin{bmatrix} -K_{22} - K_{33} & K_{21} & K_{31} \\ K_{12} & -K_{11} - K_{33} & K_{32} \\ K_{13} & K_{23} & -K_{11} - K_{22} \end{bmatrix}$$

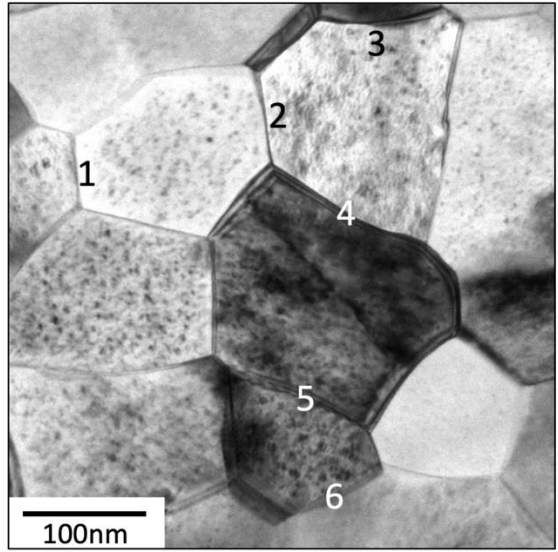

**Fig. S1.**
Six GBs attached to three grains of interest were extracted from the Figure 1(b) for GB character analysis. The GB misorientation and the GB normal are listed in Table S1.

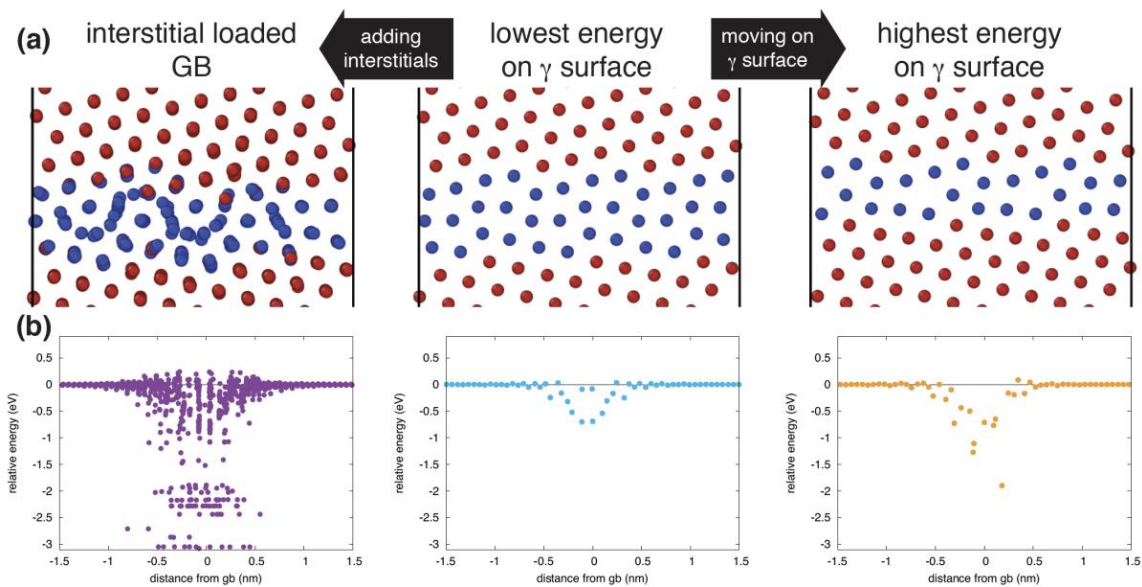

**Fig. S2.**

Role of microstate on defect interactions. (a, middle) The structure of the S13 [111] (134) symmetric tilt GB as determined by mapping the gamma surface and identifying the lowest energy structure. The microstate was modified in two ways: (right) by choosing the highest energy minimum on the gamma surface and (left) by adding ten interstitials to the lowest energy structure. Red atoms indicate Fe atoms with BCC coordination, as identified by OVITO [1], while blue atoms indicate atoms with other coordination. (b) The relative energy of vacancies at and near the three grain boundary structures, highlight the significant impact of the boundary microstate on the vacancy energetics. Each panel in (b) corresponds to the panel in (a) directly above.

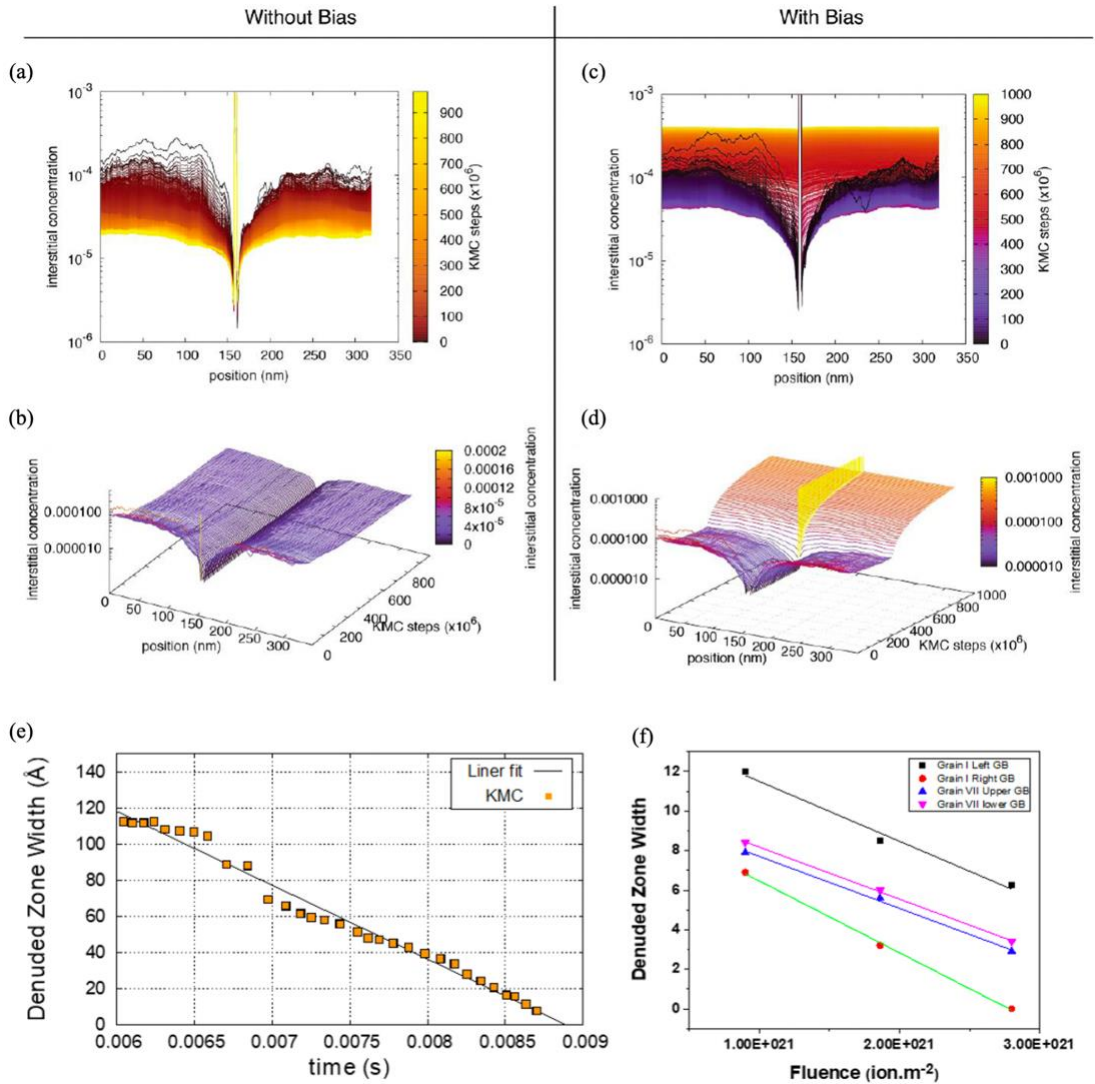

**Fig. S3.**
Time evolution (measured in kMC steps) of the interstitial concentration profiles near a grain boundary. (a-b) without and (c-d) with a bulk bias for vacancies. The results shown in (a) and (b) are identical; they are just rendered differently for clarity. The same is true of the results in (c) and (d). The concentrations at the grain boundary plane itself are clipped in some of the figures for clarity. (e) Average denuded zone width as found from the kMC simulations for conditions in which a bulk bias exists. The width is an average of the two values measured on each side of the grain boundary plane. The time over which this width is measured is focused on the time in which the profiles transition from the quasi-steady state region highlighted in blue in (d) to the purple/red region in which the interstitial concentration profiles are flat. (f) denuded zone width near-GB region in Grain I and VII (labeled in Fig. 1(a)) as a function of irradiation fluence is experimentally determined.

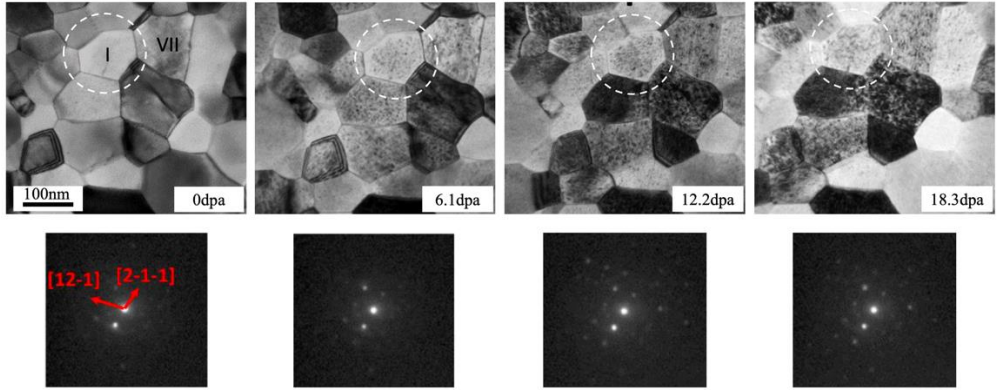

**Fig. S4**.
Sequence of TEM images taken throughout 10 keV He$^+$ irradiation, with corresponding diffraction information for each image. Certain boundaries were monitored throughout the irradiation process. For the right-hand GB in grain I, a DZ develops at 6.1 peak dpa (fluence of $0.93 \times 10^{21}$m$^{-2}$), is clearly present at 12.1 peak dpa (fluence of $1.86 \times 10^{21}$m$^{-2}$), and then disappears by a dose of 18.3 peak dpa (fluence of $2.8 \times 10^{21}$m$^{-2}$). The figure is reproduced from [2].

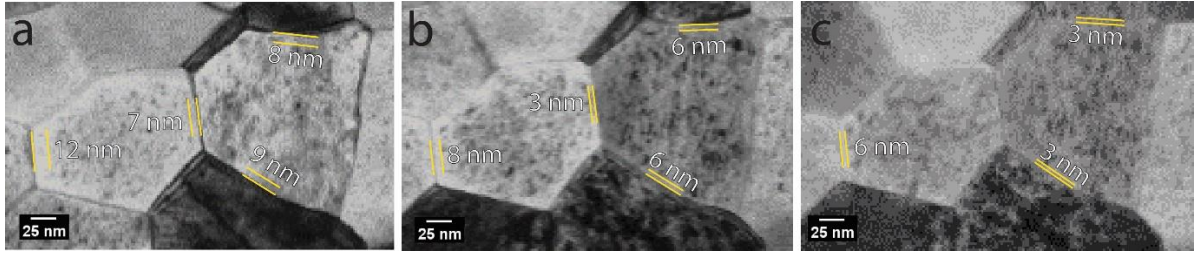

**Fig. S5**.
Sequence of bright field TEM images taken throughout 10 keV He$^+$ irradiation (a-c) at 6.1, 12.2, and 18.3 dpa, respectively. Each image includes denuded zones indication as noted in Fig. S3(f) for the Grain I left/right GBs, and Grain VII top/bottom GBs. The grain I right GB is not labeled in (c), indicating denuded zone collapse.

| GB | Misorientation angle | Rotation Axis | GB plane normal |
|---|---|---|---|
| 1 | 44.1° | $[8\,\bar{9}\,\bar{7}]$ | $[\bar{3},13,1]_A$ ; $[\bar{18},0,10]_B$ |
| 2 | 39° | $[12,\overline{15},5]$ | $[26,1,\overline{14}]_A$ ; $[22,72,\bar{5}]_B$ |
| 3 | 52.7° | $[1,\bar{3},\bar{2}]$ | Curved GB |
| 4 | 47.1° | $[21,12,\bar{1}]$ | $[\overline{53},\overline{24},10]_A$ ; $[14,\overline{22},5]_B$ |
| 5 | 55.5° | $[21,\overline{22},\bar{1}]$ | $[9,\overline{16},5]_A$; $[\overline{13},5,1]_B$ |
| 6 | 42.2° | $[4,1,\bar{4}]$ | $[\overline{12},\bar{8},1]_A$; $[1,\overline{10},0]_B$ |

**Table S1.**

Misorientation (angle/ axis) and GB plane normal for GBs labeled in Fig. S1 are listed. Analysis method is based on ref [3].

**Supplementary Information References**
1. A. Stukowski, Visualization and analysis of atomistic simulation data with OVITO–the Open Visualization Tool, Model. Simul. Mater. Sci. Eng. 18 15012 (2009).
2. O. El-Atwani, J.E. Nathaniel, A.C. Leff, K. Hattar, M.L. Taheri, Direct Observation of Sink-Dependent Defect Evolution in Nanocrystalline Iron under Irradiation, Sci. Rep. 7 1836 (2017).
3. A.C. Leff, B. Runnels, A. Nye, I.J. Beyerlein, M.L. Taheri, Determination of minimal energy facet structures in Σ3 and Σ9 grain boundaries: Experiment and simulation, Materialia. 5 100221 (2019).
4. P. Moeck, S. Rouvimov, E.F. Rauch, M. Véron, H. Kirmse, I. Häusler, W. Neumann, D. Bultreys, Y. Maniette, S. Nicolopoulos, High spatial resolution semi-automatic crystallite orientation and phase mapping of nanocrystals in transmission electron microscopes, Cryst. Res. Technol. 46 589–606 (2011).
5. J. Portillo, E.F. Rauch, S. Nicolopoulos, M. Gemmi, D. Bultreys, Precession Electron Diffraction Assisted Orientation Mapping in the Transmission Electron Microscope, Mater. Sci. Forum. 644 1–7 (2010).
6. J.F. Nye, Some geometrical relations in dislocated crystals, Acta Metall. 1 153–162 (1953).
7. W.M. Lai, D.H. Rubin, D. Rubin, E. Krempl, Introduction to Continuum Mechanics, Elsevier, .
8. A.C. Leff, C.R. Weinberger, M.L. Taheri, Estimation of dislocation density from precession electron diffraction data using the Nye tensor, Ultramicroscopy. 153 9–21 (2015).